# Rational Mechanics Theory of Turbulence

Xiao Jianhua

Henan Polytechnic University, Jiaozuo City, Henan, P.R.C., 454000

**Abstract:** The instant Lagranian coordinator system is used to describe the fluid material motion. By this way, the instant deformation gradient (expressed by spatial velocity gradient) concept is established. Based on this geometrical understanding, the strain rate and stress is expressed by local rotation tensor which is simply based on the fluid material kinetic energy concept (ruling out the concept of fluid-stretching). For fluid filling-in experiments, the characteristic of turbulence is analyzed geometrically and, then, the Navier-Stokes equation is used to get the analytical solution in first-order approximation. For convenient the comprising with experiments, the related typical values are given. As this solution is very basic, it can be expected be valuable for industrial application.

## 1. Introduction

The turbulence phenomenon is heavily related with low density fluid (such as water) and /or low-viscosity fluid (such as air). Traditionally, the velocity gradient is used to describe the motion of fluid as continuum [1]. Using strain rate and Navier-Stokes equation, its effectiveness for non-turbulence flow is widely accepted. However, for turbulence, very hard to say it can give out first-order approximation even for the simple fluid filling-in experiment which is mentioned in many text books and is observable in daily life.

Lodge [2] points out that the geometrical description of fluid material instant motion is very important. He tried to use the body-tensor to expresses the fluid motion. However, for Lagranian coordinators, tracking the fluid material become non-possible for turbulence. Even in color fluid filling-n experiment, the mixture feature of fluid material will make the Lagranian coordinator system be indefinable.

As it is known that the velocity gradient will determine an instant configuration deformation, the continuum mechanics method is widely used to attack the turbulence problem. For fluid, on intrinsic sense, the stretching like in elastic material is not acceptable. Therefore, many statistic models [3] are developed where the fluid material is viewed as particles. However, in experiments the symmetrical strain rate $e_{ij}$ indeed can be calculated and the stress can be tested. Therefore, it is identified that the turbulence is related with the vorticity in some intrinsic ways [4-5]. This interpretation, in fact, rules out the stretching concept for turbulence although some researchers tried to save the stretching concept as the symmetrical strain is always accompanied with turbulence. This contradict is very sharp in very-low viscosity fluid.

This geometrical problem is solved in this research based on Chen Zhida's S+R decomposition [6-7] for deformation tensor. In Chen's theory, the S represents intrinsic stretching. For very-low viscosity fluid, it is zero. Then the R (orthogonal rotation) is the unique tensor to describe the intrinsic motion of fluid material. According to his theory, the R is completely defined by the vorticity. The most striking feature is that the S=0 will logically require that the symmetrical strain rate must exist and can be completely expressed by vorticity. As the vorticity is related with the anti-symmetrical velocity gradient tensor component, therefore such a theory is still in continuum mechanics range. As it is seen in this paper, the intrinsic reason for the symmetrical strain rate is the material local rotation it has no relation with the elastic stretching for very-low viscosity fluid. Rather, it is the non-stretching feature makes the local rotation must form the symmetrical strain rate. Therefore, the symmetrical strain rate for non-stretchable fluid is well explained by local rotation in this research.

The century efforts to solve turbulence problem have raised the doubt whether the Navier-Stokes equation can solve the turbulence problem [8]. This paper will not address this problem. After the geometrical problem is solved, the research directly uses the Navier-Stokes equation to get the first-order analytical solution. The results can well explain the fluid filling-in phenomenon with testable filling-in parameters and fluid feature. Sure, strict exact experiments are waiting to be done in future to check the accuracy of the solutions.

The paper will firstly define the motion of fluid material with R geometry. Secondly, the geometry is used to expresses the classical symmetrical strain rate and stress. After the mechanics description about the characteristic of turbulence for fluid-filling-in experiments results, the paper goes to use the Navier-Stokes equation to establish the motion equation and get its analytical solution with the boundary condition at filling-in hole position.

## 2. Basic Definition Equations for Very-low Viscosity Fluid Motion

For an element of fluid material, it is described by inertia velocity field $\vec{u}(x^1, x^2, x^3, t)$ where the inertia coordinator $x^i$ is appropriately selected, usually $x^i$ is named as spatial coordinators. For a reference time $t_0$, the gauge tensor $g_{ij}(t_0)$ of the inertia coordinator $x^i$ system can be selected as $g_{ij}^0$. The inertia coordinator can be defined as instant Lagranian coordinator at reference time $t_0$. By this way, the same material element can be identified by the same coordinator at a tater time $t = t_0 + \Delta t$. As the fluid material can have deformation during time interval $\Delta t$, the gauge tensor of Lagranian coordinator will changed into $g_{ij}(t_0 + \Delta t)$. The instant Lagranian coordinator and the gauge tensor $g_{ij}(t_0 + \Delta t)$ define the instant commoving coordinator system (later will referred as instant dragging coordinator system [9]).

It is widely accepted that, referring to initial configuration $g_{ij}^0$, the gauge tensor of current configuration can be expressed as:

$$g_{ij} = F_i^l F_j^k g_{lk}^0 \tag{1}$$

Where, the rate deformation gradient tensor $F_j^i$ is defined as:

$$F_j^i = \delta_j^i + \frac{\partial u^i}{\partial x^j} \tag{2}$$

By this method, during time interval $\Delta t$, the spatial distance variation components of two material elements with spatial distance components $\Delta x^i$ at reference time is expressed in the instant dragging

coordinator system, as:

$$[u^i(x+\Delta x) - u^i(x)]\Delta t = \frac{\partial u^i(x)}{\partial x^j}\Delta x^j \Delta t \tag{3}$$

The current spatial distance components $\Delta X^i$ can be expressed as:

$$\Delta X^i = [u^i(x+\Delta x) - u^i(x)]\Delta t + \Delta x^i = (\frac{\partial u^i(x)}{\partial x^j}\Delta t + \delta^i_j)\Delta x^j \tag{4}$$

For unit time interval $\Delta t = 1$, denoting the instant base vector in reference time configuration as $\vec{g}_i^0$ and the instant base vector in current configuration as $\vec{g}_i$, the above equation implies that:

$$\vec{g}_i(x,t) = F_i^{\ j}(x,t)\vec{g}_j^0(x,t) \tag{5}$$

Therefore, the rate deformation gradient tensor $F_j^i$ is interpreted as the unit-time base vector transformation tensor.

For simplicity in this paper, the standard rectangular coordinator system is selected as the spatial coordinator system. Then, the Lagranian coordinator system at reference time $t_0$ is always selected as standard rectangular coordinator system. That means $g_{ij}^0 = \delta_{ij}$. By this way, the instant dragging coordinator system is always renewed with fluid configuration at reference time. By this selection, the Equation (5) is simplified as:

$$\vec{g}_i(x,t) = F_i^{\ j}(x,t)\vec{g}_j^0 \tag{6}$$

Here, $\vec{g}_i^0$ is identical with the base vector of standard rectangular coordinator system (here the spatial coordinator system).

On the other hand, in the instant dragging coordinator system, the transient kinetic energy of unit-volume material element is known as:

$$W(x, t_0 + \Delta t) = \rho(x)g_{ij}(x, t_0 + \Delta t)u^i(x, t_0)u^j(x, t_0) \tag{7}$$

Where, tensor $g_{ij}$ is the gauge tensor of instant dragging coordinator system, $\rho$ is mass density.

Without generality loss, taking the time interval be unit, for the unit-volume material element, unit-time kinetic energy variation related with fluid deformation can be expressed as:

$$\Delta W(x,t_0) = \rho(x)[g_{ij}(x,t_0) - \delta_{ij}]u^i(x,t_0)u^j(x,t_0) \tag{8}$$

Using Equation (1), it can be rewritten as unit-time unit-mass kinetic energy variation:

$$\Delta W / \rho = g_{ij}u^i u^j - u^i u^i = (F_i^l u^i)(F_j^l u^j) - u^i u^i = U^l U^l - u^i u^i \tag{9}$$

Hence, a new velocity $U^i$ can be introduced as:

$$U^i(x,t) = F_j^i(x,t)u^j(x,t) \tag{10}$$

From geometrical consideration, the spatial coordinator $x$, at unit time late, will be redefined as the instant Lagranian coordinators. As at a unit time late, the spatial coordinators $x$ will be taken by new material element, it is reasonable to name the velocity $U^i$ as instant Lagranian velocity to distinguish from the inertia velocity of the new material element. When the unit time interval is small enough for the fluid motion under discussion, as a limit, the differences between the actual position of the material element and the spatial coordinators position can be ignored. In this sense, the instant Lagranian velocity can be defined in the spatial coordinators system. Thus, in each spatial position, there are two velocity fields: inertia velocity and Lagranian velocity.

By this way, the related calculation can be taken at the spatial coordinators with the understanding that the material indifference principle is maintained.

Observing Equations (9) and (10), according to geometrical consideration related, the unit-time base vector transformation tensor must be orthogonal tensor. On the other hand, such an orthogonal tensor must be expressed by inertia velocity field. Such an orthogonal tensor does exist. This will be discussed in next section.

## 3. Strain Rate and Stress Tensor for Turbulence

Observing Equations (9) and (10), there are three possible orthogonal rotations which correspond to positive, zero, and negative unit-time unit-mass kinetic energy variations respectively. As the orthogonal expansion and contraction corresponds to a special case for this paper, it will be discussed a little later.

### 3.1. Positive or Negative unit-time unit-mass kinetic energy variations

For non-zero unit-time unit-mass kinetic energy variations, (ruling out the anisotropic orthogonal expansion) the orthogonal rotation tensor can be written as:

$$F_j^i = \frac{1}{\cos\theta} \tilde{R}_j^i \tag{11}$$

Based on related research (Chen ZD and Xiao JH [10]), for arbitral velocity gradient, the following decomposition exists:

$$F_j^i = \frac{\partial u^i}{\partial x^j} + \delta_j^i = \tilde{S}_j^i + \frac{1}{\cos\theta}\tilde{R}_j^i \tag{12}$$

Where:

$$\tilde{S}_j^i = \frac{1}{2}(\frac{\partial u^i}{\partial x^j} + \frac{\partial u^j}{\partial x^i}) - (\frac{1}{\cos\theta} - 1)(\tilde{L}_k^i \tilde{L}_j^k + \delta_j^i) \tag{13}$$

$$(\cos\theta)^{-1}\tilde{R}_j^i = \delta_j^i + \frac{\sin\theta}{\cos\theta}\tilde{L}_j^i + (\frac{1}{\cos\theta} - 1)(\tilde{L}_k^i \tilde{L}_j^k + \delta_j^i) \tag{14}$$

$$\tilde{R}_j^i = \delta_j^i + \sin\theta \cdot \tilde{L}_j^i + (1-\cos\theta)\tilde{L}_k^i \tilde{L}_j^k \tag{15}$$

$$\tilde{L}_j^i = \frac{\cos\theta}{2\sin\theta}(\frac{\partial u^i}{\partial x^j} - \frac{\partial u^i}{\partial x^j}^T) \tag{16}$$

$$(\cos\theta)^{-2} = 1 + \frac{1}{4}[(\frac{\partial u^1}{\partial x^2} - \frac{\partial u^2}{\partial x^1})^2 + (\frac{\partial u^2}{\partial x^3} - \frac{\partial u^3}{\partial x^2})^2 + (\frac{\partial u^3}{\partial x^1} - \frac{\partial u^1}{\partial x^3})^2] \tag{17}$$

The parameter $\theta$ represents local average rotation and its value range is $(-\frac{\pi}{2}, \frac{\pi}{2})$, $\tilde{L}_j^i$ represents the local average rotation direction tensor, $\tilde{R}_j^i$ is an unit-orthogonal rotation tensor, $\tilde{S}_j^i$ is a symmetric tensor which expresses the intrinsic stretching. Note that $\tilde{L}_j^i$ only has three independent parameters.

In fact, for very-low viscosity fluid material, the material element has very little intrinsic stretching ability, therefore, it is reasonable to let $\tilde{S}_j^i = 0$. Further more, for incompressible fluid without bubble, the isotropic or orthogonal expansion is not possible. In this case, the classical strain rate $e_{ij}$ is:

$$e_{ij} = \frac{1}{2}(\frac{\partial u^i}{\partial x^j} + \frac{\partial u^j}{\partial x^i}) = (\frac{1}{\cos\theta} - 1)(\tilde{L}_k^i \tilde{L}_j^k + \delta_j^i) \qquad (18)$$

It means that the classical strain rate is completely determined by the curl of velocity field.

For very-low viscosity $\mu$ fluid, the classical stress is:

$$\sigma_{ij} = -p_0 \delta_{ij} + 2\mu e_{ij} \qquad (19)$$

Based on statistic physic interpretation, the static pressure $p_0$ of fluid is caused by the dynamic motion of material elements with inertia velocity $V_0$. The static pressure can be absorbed by the homogenous distribution of rotation direction on statistic sense, by letting:

$$p_0 = 2\mu(\frac{1}{\cos\theta_0} - 1) \qquad (20)$$

In this way, the static pressure $p_0$ of fluid is interpreted as the random rotation of fluid material element with mean rotation angular $\theta_0$ and the homogenous distribution of rotation direction (note that the static inertia mean velocity $V_0$ relates to static fluid temperature and is not absorbed by static pressure). Hence, the classical stress can be re-expressed as:

$$\sigma_{ij} = 2\mu(\frac{1}{\cos\theta} - 1)\tilde{L}_i \tilde{L}_j - p_0 \delta_{ij} \qquad (21\text{-}1)$$

$$\sigma_{ij} = 2\mu[(\frac{1}{\cos\theta} - 1)\tilde{L}_i \tilde{L}_j - (\frac{1}{\cos\theta_0} - 1)\delta_{ij}] \qquad (21\text{-}2)$$

Here, the $\tilde{L}_i$ is the rotation direction unit vector components. In fact, by Equations (16) and (17), let:

$$\tilde{L}_i = e_{ijk}\tilde{L}_k^j \qquad (22)$$

Here, $e_{ijk}$ is ordering skew tensor. One will have:

$$(\tilde{L}_1)^2 + (\tilde{L}_2)^2 + (\tilde{L}_3)^2 = 1 \qquad (23\text{-}1)$$

$$\tilde{L}_k^i \tilde{L}_j^k + \delta_j^i = \tilde{L}_i \tilde{L}_j \qquad (23\text{-}2)$$

The Equations (18) and (21) give out the strain rate and stress tensors.

Further more, the unit-time unit-mass kinetic energy variation is:

$$\Delta W / \rho = [(\frac{1}{\cos\theta})^2 - (\frac{1}{\cos\theta_0})^2](V_0)^2 \qquad (24)$$

Where, the static inertia mean velocity $V_0$ of static fluid can be determined by static fluid temperature and static pressure (or $\theta_0$). The positive unit-time unit-mass kinetic energy variations will require that $\theta > \theta_0$. The negative unit-time unit-mass kinetic energy variations will require that $\theta < \theta_0$.

**3.2. Zero unit-time unit-mass kinetic energy variations**

For zero unit-time unit-mass kinetic energy variations, the orthogonal rotation tensor can be written as:

$$F_j^i = \frac{\partial u^i}{\partial x^j} + \delta_j^i = R_j^i \qquad (25)$$

Based on related research (Chen ZD [9]), for arbitral velocity gradient, when the condition:

$$\frac{1}{4}[(\frac{\partial u^1}{\partial x^2} - \frac{\partial u^2}{\partial x^1})^2 + (\frac{\partial u^2}{\partial x^3} - \frac{\partial u^3}{\partial x^2})^2 + (\frac{\partial u^3}{\partial x^1} - \frac{\partial u^1}{\partial x^3})^2] < 1 \qquad (26)$$

Is met, the following decomposition exists:

$$F^i_j = \frac{\partial u^i}{\partial x^j} + \delta^i_j = S^i_j + R^i_j \tag{27}$$

Where,

$$S^i_j = \frac{1}{2}(\frac{\partial u^i}{\partial x^j} + \frac{\partial u^j}{\partial x^i}) - (1-\cos\Theta)L^i_k L^k_j \tag{28}$$

$$R^i_j = \delta^i_j + \sin\Theta \cdot L^i_j + (1-\cos\Theta)L^i_k L^k_j \tag{29}$$

$$L^i_j = \frac{1}{2\sin\Theta}(\frac{\partial u^i}{\partial x^j} - \frac{\partial u^j}{\partial x^i}) \tag{30}$$

$$(\sin\Theta)^2 = \frac{1}{4}[(\frac{\partial u^1}{\partial x^2} - \frac{\partial u^2}{\partial x^1})^2 + (\frac{\partial u^2}{\partial x^3} - \frac{\partial u^3}{\partial x^2})^2 + (\frac{\partial u^3}{\partial x^1} - \frac{\partial u^1}{\partial x^3})^2] \tag{31}$$

The parameter $\Theta$ represents local average rotation and its value range is $(-\frac{\pi}{2}, \frac{\pi}{2})$, $L^i_j$ represents the local average rotation direction tensor, $R^i_j$ is an unit-orthogonal rotation tensor.

With the similar consideration, strain rate tensor is:

$$e_{ij} = \frac{1}{2}(\frac{\partial u^i}{\partial x^j} + \frac{\partial u^j}{\partial x^i}) = (1-\cos\Theta)L^i_k L^k_j = (1-\cos\Theta)(L_i L_j - \delta_{ij}) \tag{32}$$

To absorb the static pressure, mean rotation angular $\Theta_0$ can be introduced:

$$p_0 = 2\mu(1-\cos\Theta_0) \tag{33}$$

The stress tensor is:

$$\sigma_{ij} = 2\mu(1-\cos\Theta)(L_i L_j - \delta_{ij}) - p_0 \delta_{ij} \tag{34-1}$$

$$\sigma_{ij} = 2\mu[(1-\cos\Theta)(L_i L_j - \delta_{ij}) - (1-\cos\Theta_0)\delta_{ij}] \tag{34-2}$$

Here, the $L_i$ is the rotation direction unit vector components, which are defined as: $L_i = e_{ijk}L^j_k$. Such an orthogonal rotation is different from positive unit-time unit-mass kinetic energy variations case, in that the rotation angular and rotation direction are differently defined.

Further more, the zero unit-time unit-mass kinetic energy variation is:

$$\Delta W / \rho = [(\frac{1}{\cos\Theta})^2 - (\frac{1}{\cos\Theta_0})^2](V_0)^2 \tag{35}$$

The zero unit-time unit-mass kinetic energy variations will require that $\Theta = \Theta_0$. It means that the fluid material element rotation is changed from random to deterministic.

### 3.3. Anisotropic orthogonal expansion

For anisotropic orthogonal expansion, the orthogonal tensor only can be written as:

$$F^i_j = \begin{vmatrix} 1+\varepsilon_{11} & 0 & 0 \\ 0 & 1+\varepsilon_{22} & 0 \\ 0 & 0 & 1+\varepsilon_{33} \end{vmatrix} \tag{36}$$

According to the Equation (2), the strain rate is:

$$\varepsilon_{(ii)} = \frac{\partial u^i}{\partial x^i} \tag{37}$$

The small brackets show no summation over the index. To meet Equation (36), the following condition must be met:

$$\varepsilon_{ij} = \frac{1}{2}(\frac{\partial u^i}{\partial x^j} + \frac{\partial u^j}{\partial x^i}) = 0, \text{ for } i \neq j \qquad (38\text{-}1)$$

$$w_{ij} = \frac{1}{2}(\frac{\partial u^i}{\partial x^j} - \frac{\partial u^j}{\partial x^i}) = 0 \qquad (38\text{-}2)$$

Clearly, it corresponds to traditional elastic deformation. It has no turbulent feature. The stress is:
$$\sigma_{ij} = -p_0 \delta_{ij} + 2\mu(\varepsilon_{11} + \varepsilon_{22} + \varepsilon_{33})\delta_{ij} \qquad (39)$$

For incompressible fluid without bubble, according to definition of incompressibility, one has:
$$(\varepsilon_{11} + \varepsilon_{22} + \varepsilon_{33}) = 0 \qquad (40)$$

The stress is simply the static stress, therefore such a motion mode can be ruled out for turbulence discussion in this paper. For compressible fluid or incompressible fluid with bubble, its contribution can be absorbed into the static pressure (or more exactly, dynamic pressure).

Summering above research results, for incompressible fluid, only two orthogonal rotations (Equation (11) and Equation (25)) can be related with turbulence. In the following section, this theory will be used to explain a typical turbulence phenomenon.

## 4. Characteristics of Turbulence Structure for Simple Filling Mode

Considering a kind of color fluid is filling into an infinite tank containing very-low viscosity fluid through a small hole (see Fig. 1). When the filling velocity is high enough, turbulence flow will be formed, which can be observed by the color fluid motion. For simplicity, the color fluid is supposed to be the same kind of fluid with the fluid in tank but colored. The filling direction (the normal of filling side) is taken as $x^3 = z$, the other two coordinators are selected as shown in Fig.1. The zero point is selected at the center of filling hole.

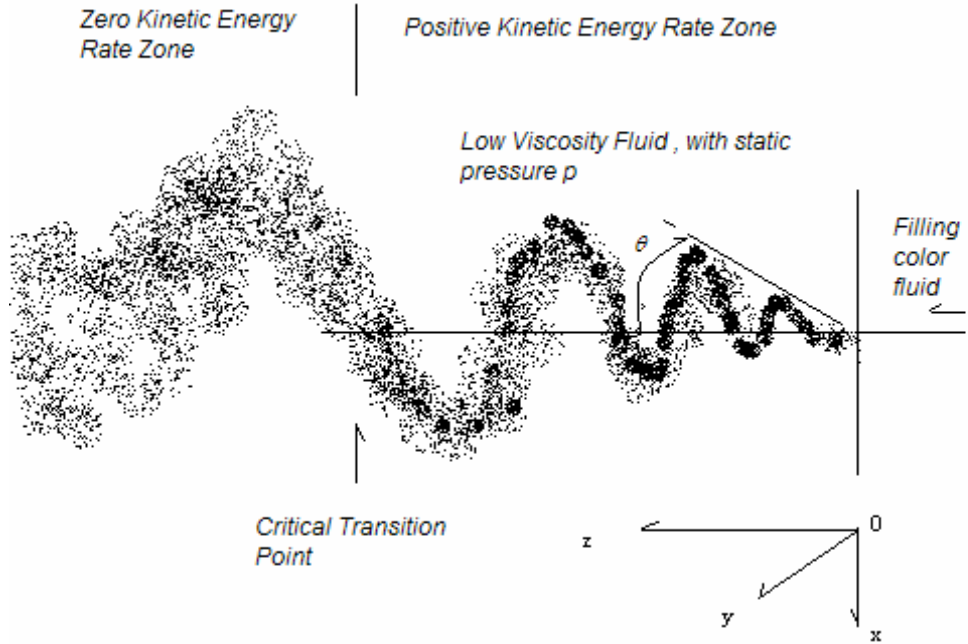

Fig.1. Turbulence Produced by Simple Filling Mode

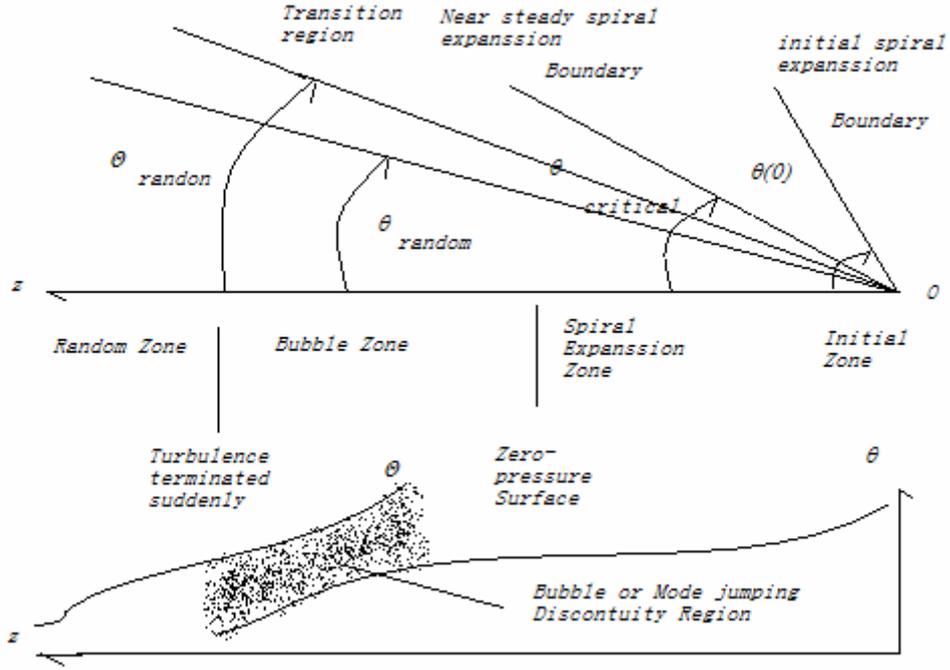

*Fig.2. Turbulence Structure:Rotation Angular*

### 4.1. Orthogonal Rotation at the Hole Position

At the color fluid filling hole position, the fluid stress $\sigma_{ij}$ must equate the filling color fluid pressure $P_{in}$ (for filling the color fluid into the tank, $P_{in} > p_0 > 0$ ). Without considering the effects of hole size and shape, the boundary condition is:

$$\sigma_{33} = 2\mu(\frac{1}{\cos\theta} - 1)\tilde{L}_3\tilde{L}_3 - p_0 = P_{in}, \text{ at } x = y = z = 0 \tag{41-1}$$

The other stress components are static pressure (that means the side effects are omitted).

$$\sigma_{ij} = 2\mu(\frac{1}{\cos\theta} - 1)\tilde{L}_i\tilde{L}_j - p_0\delta_{ij} = -p_0, \text{ at } x = y = z = 0 \tag{41-2}$$

It shows that, at the hole position, the fluid motion is described orthogonal rotation:

$$\tilde{L}_3 = 1, \quad \tilde{L}_1 = \tilde{L}_2 = 0 \tag{42-1}$$

$$\theta(0) = \arccos[\frac{1}{1 + (P_{in} + p_0)/(2\mu)}] \tag{42-2}$$

The rotation angular $\theta(0)$ shows that the color fluid will rotate around the filling-in direction with angular $\theta(0)$ while moving along the filling-in direction. This forms the initial turbulence pattern. Outside the angular $\theta(0)$, the color fluid will not come in. For very-low viscosity fluid or very high filling-in pressure, the angular $\theta(0)$ tends to $\frac{\pi}{2}$.

Based on Equations (18) and (24), the only non-zero symmetric strain rate component is:

$$e_{33} = \frac{\partial u^3}{\partial x^3} = (\frac{1}{\cos\theta} - 1)\tilde{L}_3\tilde{L}_3 = \frac{P_{in} + p_0}{2\mu} \tag{43-1}$$

It shows that at the hole position, the fluid inertia velocity field is increased along the $z = x^3$ direction

(that is the filling-in direction). Thus a positive unit-mass unit-time kinetic energy variation zone is formed ahead of the hole. The local volume expansion rate is equal to the color fluid filling-in volume rate:

$$k(0) = e_{33} = (\frac{1}{\cos\theta} - 1) = \frac{P_{in} + p_0}{2\mu} \tag{43-2}$$

Watching from $z$ direction, the color fluid trace is spiral expansion line (see Fig.1). Watching from $x$ (or $y$) direction, the color fluid trace is on a cone surface determined by angular $\theta(0)$ (see Fig.1).

By Equations (10) and (42-2), the fluid Lagranian velocity absolute value for original static fluid material (non-color material) is:

$$U(0) = \frac{1}{\cos\theta(0)} V_0 = (1 + \frac{P_{in} + p_0}{2\mu}) V_0 \tag{44-1}$$

The striking feature is the fluid velocity not only is determined by the filling-in pressure but also the static fluid parameters. As the $V_0$ is related with temperature, the temperature dependence is very striking. It shows that to make good observation in turbulent experiments the related parameters should be carefully selected.

For color fluid material, the Lagranian velocity absolute value is:

$$\tilde{U}(0) = \frac{1}{\cos\theta(0)} V_{in} = (1 + \frac{P_{in} + p_0}{2\mu}) V_{in} \tag{44-2}$$

Where, the $V_{in}$ is the initial velocity of color fluid material. The acceleration phenomenon is very clear.

Note that material trace is a spiral line rotating along $z$ direction, hence the progressive speed along $z$ direction for hole position will depend on $d\theta(0,t)/dt$. This quantity should be determined by motion equations. That will be discussed later.

### 4.2. Orthogonal Rotation at the Positive Kinetic Energy Variation Zone

For static low viscosity fluid with static pressure $p_0$, when the color fluid is filling into the tank, the material element near the filling hole will have positive unit-mass unit-time kinetic energy variation. As (at the hole position) the rotation is along the filling-in direction, it can expect that near the hole position the rotation will divert from this direction. Because along the rotation direction the orthogonal rotation gives out a positive stress component contribution, there must exist a zero stress surface. This surface is defined by the equation:

$$\sigma_{ij} = 2\mu(\frac{1}{\cos\theta} - 1)\tilde{L}_i\tilde{L}_j - p_0\delta_{ij} = 0 \tag{45-1}$$

$$e_{ij} = (\frac{1}{\cos\theta} - 1)\tilde{L}_i\tilde{L}_j \tag{45-2}$$

The Equation (45-2) shows that the principle direction of strain rate is just the rotation direction. Only on the rotation direction, the equations have a solution. Using the rotation direction as the reference frame, the stress and strain rate is:

$$\sigma_{LL} = 0, \text{ other is } -p_0 \tag{46-1}$$

$$e_{LL} = \frac{1}{\cos\theta_{critical}} - 1 = \frac{p_0}{2\mu}, \text{ other is zero} \tag{46-2}$$

The Equation (45-1) gives out the rotation angular $\theta_{critical}$ as the following:

$$\theta_{critical} = \arccos[\frac{1}{1 + p_0/(2\mu)}] \tag{46-3}$$

This angular $\theta_{critical}$ is smaller than the hole rotation angular $\theta(0)$, that means the initial cone angular $\theta(0)$ is replaced by the zero-pressure surface with cone angular $\theta_{critical}$. The angular $\theta_{critical}$ is determined by the static pressure $p_0$. On the rotation direction, the strain rate is determined to be $p_0/(2\mu)$. That means that on the zero-pressure surface the rotation direction velocity gradient is pressure dependent. For very-low viscosity fluid this velocity gradient will be very high along the rotation direction.

The volume expansion rate is:

$$k_{critical} = \frac{1}{\cos\theta_{critical}} - 1 = \frac{p_0}{2\mu} \tag{46-4}$$

Its constant feature shows that the color material is mixed with the non-color material with constant rate.

The rotation axe is defined by:

$$\tilde{L}_3 < 1, \quad (\tilde{L}_1)^2 + (\tilde{L}_2)^2 = 1 - (\tilde{L}_3)^2 \ll 1 \tag{47-1}$$

The Lagranian velocity absolute value for original static fluid material is:

$$\overline{U}_{critical} = \frac{1}{\cos\theta_{critical}} V_0 = (1 + \frac{p_0}{2\mu})V_0 \tag{47-2}$$

It means the original static fluid material will rotate along the rotation direction with this speed. It represents the velocity at the boundary between the turbulence zone and the static zone.

The Lagranian velocity absolute value for non-color material is:

$$U_{critical} = \frac{1}{\cos\theta_{critical}} U(0) = (1 + \frac{p_0}{2\mu})(1 + \frac{P_{in} + p_0}{2\mu})V_0 \tag{47-3}$$

It represents the maximum velocity in the turbulence zone for non-color fluid material.

For color fluid material, the Lagranian velocity absolute value is:

$$\tilde{U}_{critical} = \frac{1}{\cos\theta_{critical}} \tilde{U}(0) = (1 + \frac{p_0}{2\mu})(1 + \frac{P_{in} + p_0}{2\mu})V_{in} \tag{47-4}$$

It shows that once the color fluid material moving out from the hole position zone, the absolute value is increased. (It means the input energy will finely be converted into internal energy related with temperature). The difference between the color and non-color material will make the unit-volume fluid form micro-scope turbulence. However, this problem will not be addressed here.

Therefore, along the rotation axe $(L_1, L_2, L_3)$, the color fluid will rotate with angular $\theta_{critical}$ along the axe direction with moving speed $\tilde{U}_{critical}$, while the non-color fluid material rotate with angular $\theta_{critical}$ along the axe direction with moving speed ranging from $\overline{U}_{critical}$ to $U_{critical}$. Note that, for this zero-pressure surface, although the rotation axe direction is diverted form $z$ direction, the rotation angular $\theta_{critical}$ is the same. When the filling-in color fluid material is very small in volume, the macro-mixed-fluid Lagranian velocity absolute value is $U_{critical}$.

As only the rotation direction strain rate is non-zero, the color fluid will travel along the rotation axe on the cone surface defined by the angular $\theta_{critical}$. Hence, the color fluid will be concentrated in the special region defined by zero-pressure surface.

Watching from $z$ direction, the color fluid trace center is spiral expansion line while the color fluid circulates around the trace center (see Fig.1).

Along spiral expansion line along the $z$ direction, between the zero-pressure trace and the hole position, the color material has the largest Lagranian velocity absolute value. After it, the non-color fluid material with Lagranian velocity absolute value ranging from $\overline{U}_{critical}$ to $U_{critical}$ will followed. They form a tail for the turbulence. This region is characterized as the speeding-up region.

In the zero-pressure region, the rotation axe direction will be further diverted, therefore the color fluid material will be diverted further. Thus, the zero-pressure region will disappear and the stress will be negative. In fact, the motion of the zero-pressure face is the turbulence center.

### 4.3. Orthogonal Rotation at the Negative Kinetic Energy Variation Zone

According to above research, the color fluid will be firstly rotate along $z$ direction, and gradually diverted from $z$ direction, then comes into the zero-pressure surface. During this process, the original static fluid is mixed with the color fluid. That makes the color fluid can transport its kinetic energy to the original static fluid.

Geometrically, when $\theta < \pi/4$, that is when:

$$\frac{1}{4}[(\frac{\partial u^1}{\partial x^2} - \frac{\partial u^2}{\partial x^1})^2 + (\frac{\partial u^2}{\partial x^3} - \frac{\partial u^3}{\partial x^2})^2 + (\frac{\partial u^3}{\partial x^1} - \frac{\partial u^1}{\partial x^3})^2] < 1 \tag{49}$$

There are two possible motion modes: $F_j^i = \frac{1}{\cos\theta}\tilde{R}_j^i$ or $F_j^i = R_j^i$. They correspond to different stress and strain rate. That is:

$$\sigma_{ij} = 2\mu(\frac{1}{\cos\theta} - 1)\tilde{L}_i\tilde{L}_j - p_0\delta_{ij} \tag{50-1}$$

$$e_{ij} = (\frac{1}{\cos\theta} - 1)\tilde{L}_i\tilde{L}_j \tag{50-2}$$

Or:

$$\sigma_{ij} = 2\mu(1 - \cos\Theta)(L_iL_j - \delta_{ij}) - p_0\delta_{ij} \tag{51-1}$$
$$e_{ij} = (1 - \cos\Theta)(L_iL_j - \delta_{ij}) \tag{51-2}$$

The continuity of stress and strain rate can never be achieved. However, in fact, the deformation is continuous. This means that what modes will take dominant position will depends on the boundary condition. To solve this problem, let's recall the related turbulent experiment like showing in Fig.1.

The related experiments show that the trace of color fluid is continuous. That means that the rotation direction must be continuous. Hence, at critical transition point, there exists:

$$\tilde{L}_i = L_i \tag{52}$$

To make the stress and strain rate be continuous, an isotropic expansion strain rate must be stacked on the $F_j^i = R_j^i$ mode. Let it be:

$$\overline{e}_{(ii)} = \frac{\partial \overline{u}^i}{\partial x^i} = k, \quad i = 1,2,3 \tag{53}$$

Then Equation (51) becomes:

$$\sigma_{ij} = 2\mu[(1 - \cos\Theta)(L_iL_j - \delta_{ij}) + k\delta_{ij}] - p_0\delta_{ij} \tag{54-1}$$
$$e_{ij} = (1 - \cos\Theta)(L_iL_j - \delta_{ij}) + k\delta_{ij} \tag{54-2}$$

Combining Equations (50), (52), and (53), the stress and strain rate continuity will require that:

$$\frac{1}{\cos\overline{\theta}} - 1 = 1 - \cos\overline{\Theta} = k \tag{55}$$

This equation defines the critical transient point. When the equation does be met by the fluid motion, it means that the transition is achieved through an isotropic expansion. For isotropic expansion, the color fluid material will go every direction.

The critical transition point unit-time deformation can be expressed as:

$$F_j^i = \frac{1}{\cos\bar{\theta}} \widetilde{R}_j^i = R_j^i + k\delta_j^i \tag{56}$$

What does mean by the critical transition point? The most striking feature is that the isotropic expansion may produce bubbles in the fluid. In fact, the bubbles are intrinsic feature for turbulence which is observed in many experiments. Under this understanding, the $k$ is the bubble production rate of unit-volume element. Putting the bubble production rate as the intrinsic feature of fluid, the critical transient point is completely determined by it. For this understanding, Equation (53) gives the bubble strain rate with the bubble velocity field $\bar{u}^i$.

For this zone, the rotation angular will jump from $\bar{\theta}$ to $\bar{\Theta}$. Comparing Equations (17) and (31), the $\bar{\Theta}$ is bigger than $\bar{\theta}$, therefore the rotation angular is jumped again (see Fig.1).

### 4.4. Random Distribution of Rotation Direction and Turbulent Decaying

Observing Equation (55), the critical value $k$ is completely determined by the curl of velocity field. The isotropic expansion will finely decay into random distribution of rotation direction. In this case, the turbulence disappears (see Fig.1).

However, if no bubble is produced, there is only one possible way to make the stress be continuous. That is to let the rotation direction $\widetilde{L}_i$ is homogenous distribution on positive $z$ direction. This can be expressed as:

$$\widetilde{L}_i = \frac{1}{\sqrt{3}} \tag{57}$$

Introducing the dynamic pressure $p$, by Equation (21-2), it can be defined as:

$$p = 2\mu(\frac{1}{\cos\theta_{random}} - 1) \tag{58-1}$$

The volume expansion rate of turbulence zone can be completely attributed to the filling-in volume rate. That is:

$$k_{random} = (\frac{1}{\cos\theta_{random}} - 1) = \frac{1}{\Omega} \cdot k(0) = \frac{1}{\Omega} \cdot \frac{P_{in} + p_0}{2\mu} \tag{58-2}$$

Where, the $\Omega$ is the total volume of the tank. Hence, the angular $\theta_{random}$ is determined by fluid feature and filling-in pressure. Therefore, although the fluid motion is random, its turbulence feature is deterministic.

In this case, to make the stress rate be continuous, for the $F_j^i = R_j^i$ mode, let the rotation direction $L_i$ is homogenous distribution on any direction, then the related strain rate is:

$$e_{ij} = -(1 - \cos\Theta_{random})\delta_{ij} \tag{59-1}$$

The stress continuity can be achieved under the condition that:

$$p = 2\mu(1 - \cos\Theta_{random}) + p_0 \tag{59-2}$$

By Equations (58-1) and (58-2), it is completely determined by fluid feature and filling-in pressure.

Once the unit-time deformation mode is changed into $F_j^i = R_j^i$ mode with homogenous distribution on any direction, the turbulence is finished. Generally, noting that temperature is proportional with kinetic energy, the fluid temperature at fully random region will be increased as the

current unit-time unit mass kinetic energy rate is (referring to Equations (20) and (24)):

$$\Delta W / \rho = [(\frac{1}{\cos\theta_{random}})^2 - (\frac{1}{\cos\theta_0})^2](V_0)^2 \tag{60-1}$$

By Equation (58-2), it is:

$$\Delta W / \rho = [(1 + \frac{P_{in} + p_0}{2\mu\Omega})^2 - (1 + \frac{p_0}{2\mu})^2](V_0)^2 \tag{60-2}$$

Summering the discussion in this section, the turbulence feature can be divided into four typical zones: (1) Initial Zone, which is deterministic speeding-up zone; (2) Spiral Expansion Zone, which is terminated by zero-pressure surface; (3) Bubble Zone, where the spiral expansion is accompanied with bubble production; (4) Random Zone, the turbulence finished suddenly. The related typical quantities are shown in Fig.2.

## 5. Turbulence Motion Equation and Solution

Now, let's turn to discus the motion equation. The motion equation can be derived from the Navier-Stokes equation as:

$$\frac{\partial \sigma_{ij}}{\partial x^j} = \rho \frac{\partial u^i}{\partial t} + \rho u^j \frac{\partial u^i}{\partial x^j} \tag{61}$$

Putting equation (21-1) into it, one finds:

$$2\mu[(\frac{1}{\cos\theta} - 1)\tilde{L}_i \frac{\partial \tilde{L}_j}{\partial x^j} + \tilde{L}_i \tilde{L}_j \frac{\partial}{\partial x^j}(\frac{1}{\cos\theta})] - \frac{\partial p_0}{\partial x^i} = \rho \frac{\partial u^i}{\partial t} + \rho u^j \frac{\partial u^i}{\partial x^j} \tag{62}$$

For our simple model the static pressure is constant. That is $\partial p_0 / \partial x^i = 0$.

### 5.1 Solution of Center Trace

Near filling-in hole position, the rotation can be simplified by the rotation along filling-in direction. At later time, the centre trace are steadily along the filling-in direction with a limit diversity. For the centre trace, the rotation direction vector components can be approximated as:

$$\tilde{L}_3 \approx 1, \quad \tilde{L}_1 = \tilde{L}_2 \approx 0, \quad \theta \approx \theta(0) \tag{63-1}$$

The velocity gradient component along the rotation direction is:

$$\frac{\partial u^3}{\partial x^3} \approx \frac{1}{\cos\theta} - 1, \text{ others zero} \tag{63-2}$$

Then, the motion Equation (62) is simplified as:

$$2\mu \frac{\partial}{\partial x^3}(\frac{1}{\cos\theta}) = \rho \frac{\partial u^3}{\partial t} + \rho u^3 \frac{\partial}{\partial x^3}(\frac{1}{\cos\theta}) \tag{64-1}$$

It can be rewritten as:

$$(2\mu - \rho u^3)\frac{\partial}{\partial x^3}(\frac{1}{\cos\theta}) = \rho \frac{\partial u^3}{\partial t} \tag{64-2}$$

As it is known that acceleration is positive for inertia velocity, the equation shows that there are two cases:

Case A $\qquad 1 - \frac{\rho u^3}{2\mu} > 0, \quad \frac{\partial}{\partial x^3}(\frac{1}{\cos\theta}) > 0 \tag{65-1}$

Case B $\qquad 1 - \frac{\rho u^3}{2\mu} < 0, \quad \frac{\partial}{\partial x^3}(\frac{1}{\cos\theta}) < 0 \tag{65-2}$

For case A, the rotation angular will increase along the rotation direction so rapidly decay into random distribution and motion. This will not produce macro-observable turbulence, so will not be studied in this paper.

For case B, the rotation angular will decrease along the rotation direction. As Reynolds number is defined as:

$$R_e = \frac{l \cdot u^3}{2\mu} \tag{66}$$

Where, $l$ is the characteristic length scale. In fact, here the unit-volume element material is discussed so the $l$ is unit. For high Reynolds number, the mode must be case B.

For very high Reynolds number (very-low viscosity fluid), the Equation (65-2) can be approximated as:

$$\frac{\partial}{\partial x^3}(\frac{1}{\cos\theta}) = -\frac{1}{u^3} \cdot \frac{\partial u^3}{\partial t} \tag{67}$$

At hole point, the boundary condition is:

$$\theta(x^3, t) = \theta(0), \text{ at } x^3 = 0, \text{ for } t > 0 \tag{68-1}$$

$$\frac{\partial u^3(x^3, t)}{\partial t} = \frac{P_{in} - p_0}{\rho}\Pi, u^3 = V_{in}, \text{at } x^3 = 0, \text{ for } t > 0 \tag{68-2}$$

Where, $\Pi$ is the geometrical parameter of hole section (its idea value is 1). The solution is:

$$\frac{1}{\cos\theta} = \frac{1}{\cos\theta(0)} - \frac{P_{in} - p_0}{\rho V_{in}}\Pi \cdot z \tag{69-1}$$

Hence, combining the Equation (42-2), the rotation angular solution is:

$$\theta = \arccos[1 + \frac{P_{in} + p_0}{2\mu} - \frac{P_{in} - p_0}{\rho V_{in}}\Pi \cdot z]^{-1} \tag{69-2}$$

It shows that the rotation angular is non-linearly decreased along rotation direction. Positive $\theta$ and negative $\theta$ represent clockwise and anti-clockwise respectively, as usual definition.

By Equations (63-2) and (69-2), the spatial inertia velocity component along the rotation direction is:

$$u^3 = V_{in} + \frac{P_{in} + p_0}{2\mu} \cdot z - \frac{P_{in} - p_0}{2\rho V_{in}}\Pi \cdot z^2 \tag{69-3}$$

For small $z$, it is increased along the rotation direction. Thus, it forms a speeding-up region. The maximum velocity $u^3$ position is:

$$Z_{\max} = [\frac{P_{in} + p_0}{2\mu}]/[\frac{P_{in} - p_0}{2\rho V_{in}}\Pi] \tag{70-1}$$

At this position, the maximum value is:

$$\text{Maximum}[u^3] = V_{in} + \frac{P_{in} + p_0}{2\mu} \cdot Z_{Max} - \frac{P_{in} - p_0}{2\rho V_{in}}\Pi \cdot Z_{Max}^2 \tag{70-2}$$

This position defines a plane where the symmetric stress is identical with the static pressure. So, it defines the end position of turbulence. Surely, the pre-condition is that the near-hole solution keeps a good approximation, which depends on the filling-in parameters and fluid feature. After this position, the strain rate $e_{33}$ will be negative. The negative sign shows that the rotation direction will be diverted from the $z$ direction. Therefore, it becomes the conventional fluid flow.

The other two spatial velocity components can be determined by solving definition equations with suitable boundary condition. However, for this paper, this solution is not required. So, they will be not discussed.

### 5.2 Solution Near Centre Trace Region

When the color material spirally expanding and progressively move along the rotation direction, the rotation direction will be diverted from the $z$ direction a little. This can be approximated as:

$$\tilde{L}_3 < 1, \quad (\tilde{L}_1)^2 + (\tilde{L}_2)^2 = 1 - (\tilde{L}_3)^2 = b < 1 \tag{71}$$

Let:

$$\frac{\partial \tilde{L}_j}{\partial x^j} = a \tag{72}$$

The parameter $a$ measures the diversity of rotation direction. For centre trace, it is zero. Hence, it measures the size of color material around the centre trace.

The motion Equation (62) can be approximated as:

$$2\mu[(\frac{1}{\cos\theta} - 1)\tilde{L}_3 \cdot a + \tilde{L}_3 \tilde{L}_j \frac{\partial}{\partial x^j}(\frac{1}{\cos\theta})] = \rho \frac{\partial u^3}{\partial t} + \rho u^j \frac{\partial u^3}{\partial x^j} \tag{73}$$

For this case, the differentiation along the rotation direction $\tilde{L}$ can be introduced as:

$$\frac{\partial}{\partial \tilde{L}}(\frac{1}{\cos\theta}) = \tilde{L}_j \frac{\partial}{\partial x^j}(\frac{1}{\cos\theta}) \tag{74}$$

Then, the Equation (73) can be rewritten as:

$$(\frac{1}{\cos\theta} - 1)a + \frac{\partial}{\partial \tilde{L}}(\frac{1}{\cos\theta}) = \frac{\rho}{2\mu\sqrt{1-b}}(\frac{\partial u^3}{\partial t} + u^j \frac{\partial u^3}{\partial x^j}) \tag{75}$$

Near the trace centre, using the Equation (64-1), it can be approximated as:

$$(\frac{1}{\cos\theta} - 1)a + \frac{\partial}{\partial \tilde{L}}(\frac{1}{\cos\theta}) \approx \frac{\rho}{2\mu\sqrt{1-b}}(\frac{\partial u^3}{\partial t} + u^3 \frac{\partial u^3}{\partial x^3}) \tag{76-1}$$

Or,

$$(\frac{1}{\cos\theta} - 1)a + \frac{\partial}{\partial \tilde{L}}(\frac{1}{\cos\theta}) \approx \frac{\rho}{\sqrt{1-b}} \frac{\partial}{\partial x^3}(\frac{1}{\cos\theta}) \tag{76-2}$$

Using the angular $\theta_{centre}$ represents the centre trace solution, the above equation becomes:

$$(\frac{1}{\cos\theta_{center}} - 1)a \approx (\frac{\rho}{\sqrt{1-b}} - 1) \cdot \frac{\partial}{\partial x^3}(\frac{1}{\cos\theta_{center}}) \tag{77}$$

By Equation (69-1), it gives out:

$$a \approx (1 - \frac{\rho}{\sqrt{1-b}}) \cdot \frac{P_{in} - p_0}{\rho V_{in}} \cdot \Pi \Big/ (\frac{1}{\cos\theta(0)} - 1 - \frac{P_{in} - p_0}{\rho V_{in}} \cdot \Pi \cdot z) \tag{78}$$

Its absolute value is increased with the $z$ coordinator increasing, as the angular $\theta_{centre}$ is decreased.

For $\rho > \sqrt{1-b}$, $a$ is negative. It does mean the diversity of rotation direction. However, for $\rho < \sqrt{1-b}$, $a$ is positive. It means the diverted rotation will concentrate toward the trace center. Therefore, the critical value of fluid mass density is unit density.

As the water has $\rho \approx 1$, for $\rho < 1$ the color trace will be very narrow and sharp. The trace is clearly observable. For $\rho > 1$, the color trace will be much wider. So, the water is very sensitive for turbulence. For $\sqrt{2} > \rho > 1$, a critical rotation direction $b_{critical}$ can be defined as:

$$b_{critical} = \rho^2 - 1 \tag{79}$$

If its value is near 1, it means the rotation is nearly on the $z = const$ plane. So, for fluid mass density

$\sqrt{2} > \rho > 1$, it defines a cone face boundary for the maximum direction of rotation axe. For $\rho \geq \sqrt{2}$, such a boundary does not exist. Hence, to make the turbulence trace be clearly viewed for fluid filling-in experiments, the mass density must be carefully selected.

Generally speaking, the solutions (78) and (69-1) describe the characteristics of turbulence in first order approximation.

At least, the approximation has enough accuracy before and near the zero-pressure surface. For bubble zone and random zone, the approximation should be replaced by other solutions. In this case, one must use the (62) to find the correct solution although it is very difficult. However, as the two modes have competition in this region, the bubble feature and other factors must be defined by other physical consideration rather than limited to Equation (62). For example, the Equation (55) must be applied for bubble case. This topic is out the range of this paper, so it will not be discussed further.

Summering up the results in this section, the turbulence centre region has first-order approximation analytic solution (that is the Equations (69-2) and (78). The related quantities can be calculated according to their corresponding equations. Hence, the conclusion is that this research gets the first-order analytical solution for turbulence produced by fluid filling-in. As this solution is very basic, it can be expected be valuable for industrial application.

## 6. Conclusion

This research shows that for the turbulence phenomenon, the geometrical description of fluid material instant motion is very important. As it is known that the velocity gradient will determine an instant configuration deformation, the fluid motion is frequently treated in continuum mechanics. For fluid, on intrinsic sense, the stretching like in elastic material is not acceptable. However, in experiments the symmetrical strain rate $e_{ij}$ indeed can be calculated and the stress can be tested. This contradict is very sharp in very-low viscosity fluid. The century efforts to solve turbulence problem have raised the doubt whether the Navier-Stokes equation can solve the turbulence problem. As it is seen in this paper, the intrinsic reason for the symmetrical strain rate is the material local rotation it has no relation with the elastic stretching for very-low viscosity fluid. Rather, it is the non-stretching feature makes the local rotation must form the symmetrical strain rate. Therefore, the symmetrical strain rate for non-stretchable fluid is well explained by local rotation in this research.

After solving contradict between the symmetrical strain rate for non-stretchable fluid, the research shows that the Navier-Stokes equation can be used to give the correct results for the turbulence. In fact, the first-order analytical solution for turbulence center region is obtained in the paper. No need to say that the results must be examined by experiments. For convenient the comprising with experiments, the related typical values are given. However, as the most available experiments do not supply the needed parameter for the view-point of the research, the author cannot finish this job. So, it is hopping that some valuable experiments be done somewhere. As this solution is very basic, it can be expected be valuable for industrial application.

**Reference**
[1] Truesdell C., *The mechanical foundations of elasticity and fluid dynamics,* Gorden & Breach Science Pub., 1966
[2] Lodge A S. Body tensor fields in continuum mechanics. Academic Press, 1974
[3] Tomas Bohr et al. Dynamical Systems Approach to Turbulence. Cambridge Uni. Pres. 1998

[4] Townsend, A.A., The Structure of Turbulent Shear Flow. Cambridge Uni. Pres. 1976

[5] Chorin, A. J., Vorticity and Turbulence. Springer, 1994

[6] Chen Zhida. Geometric Theory of Finite Deformation Mechanics for Continuum. *Mechanica Sinica,* ,No.2, 107-117, (In Chinese), 1979

[7]. Chen Zhida, Limit Rotation Expression in Non-linear Field Theory of Continuum. *Applied Mathematics and Mechanics*, 1996 No.7, 959-968, (In Chinese), 1986

[8] Foias, C., What Do the Navier-Stokes Equation Tell Us about Turbulence ?, in Harmonic Analysis and Non-linear Differential Equations, ed. M.L. Lapidus et al., Contemporary Mathematics, Vol. 208: 151-180, 1997

[9]Chen Zhida. Rational Mechanics—Non-linear Continuous Mechanics. Xuzhou: China University of Mining & Technology Press, (In Chinese) 1987

[10]Xiao Jianhua, Decomposition of displacement gradient and strain definition, in Y Luo, R Qiu ed. *Advance in Rheology and its Application (2005)*, Science Press USA Inc. 2005, 864-868